\documentclass[aps,prl,superscriptaddress,twocolumn]{revtex4}
\usepackage{bm}
\usepackage{graphicx,amsmath,latexsym,amssymb}
\usepackage{color}
\usepackage{dcolumn}

\begin{document}
\title{Violation of the action-reaction principle in an asymmetrically excited system} %: manifestation of the vacuum momentum}
\author{M. Donaire} 
\email{donaire@lkb.upmc.fr, mad37ster@gmail.com}
\affiliation{Laboratoire Kastler Brossel, UPMC-Sorbonnes Universit\'es, CNRS, ENS-PSL Research University,  Coll\`{e}ge de France, 4, place Jussieu, F-75252 Paris, France}

%\date{28 Jan 2015}

\begin{abstract}
Violation of the action-reaction principle is shown to occur in the van der Waals interaction between two atoms, one of which is excited. It is accompanied by the transfer of linear momentum to the electromagnetic vacuum. The vacuum momentum results from the asymmetric interference of the virtual photons scattered off each atom along the interatomic direction, which is in itself a manifestation of the optical theorem. This momentum, of equal strength and opposite direction to the momentum gained by the  two-atom system, is ultimately released through directional spontaneous emission. A quantitative prediction of this phenomenon is made in a two-alkali atom system. It is conjectured that action-reaction violation takes place in any asymmetrically excited system.
\end{abstract}
\maketitle

In a recent publication \cite{Me2PRA} it has been found that, in a two-atom system with one atom excited, the van der Waals (vdW) potentials of each atom differ. This implies the violation of the action-reaction principle within the system. It is the purpose of this article to explain this phenomenon theoretically, to provide a quantitative estimate for its experimental verification, and to relate it to precedent findings. The explanation lies in the fundamental laws of momentum conservation and the optical theorem, from which we conjecture that action-reaction violation applies to any asymmetrically excited system.

VdW forces between neutral atoms are the result of the quantum fluctuations of both the electromagnetic (EM) field and the atomic charges \cite{Milonnibook,Cohenbook}.  At zero temperature, in the electric dipole approximation and for two atoms in their ground states located at a distance $R$ apart,  the atoms undergo a series of virtual E1 transitions to upper levels accompained by the exchange of photons of frequency $\omega\lesssim c/R$. It is the mutual coupling of the charges of each atom to the intermediate photons that induces correlations between their transient dipole moments, giving rise to a non-vanishing interaction that can be computed within the framework of stationary quantum perturbation theory. For short interatomic distances in comparison to the relevant transition wavelengths, the resultant forces are referred to as London dispersion forces \cite{Milonnibook,Craigbook,London}. For large distances they are known as Casimir-Polder forces  \cite{Milonnibook,Craigbook,Casimir-Polder1948}.

\indent The situation is different when at least one of the atoms is  excited. In addition to the aforementioned interaction mediated by off-resonant photons, transitions to lower energy atomic levels proceed through the coupling of the charges to resonant photons. This was firstly noticed by Wylie \& Sipe \cite{Wylie}. In general, this \emph{resonant interaction} is time-dependent, since it involves a nearly reversible and periodic excitation transfer between the two atoms. For non-identical atoms, in the weak coupling regime, the probability of total excitation transfer is small, it is attenuated by the spontaneous emission of each atom in free-space and its frequency is given by the detuning between the atomic species \cite{Me2PRA,Me1PRL,Milonni,Berman}. As shown in Refs.\cite{Berman,Me1PRL,Milonni,Me2PRA,Sherkunov}, the resonant interaction is to be computed within the framework of time-dependent perturbation theory, eventhough it is quasistationary for the usual case of an adiabatic excitation \cite{Berman,Me2PRA}. Relevant to us is the fact that only the strengths of the resonant forces on each atom differ \cite{Me2PRA}. 

For simplicity, let us consider  two two-level atoms of different types, $A$ and $B$, located a distance $R$ apart, with atom $A$ excited. The transition frequencies of each atom are $\omega_{A}$ and $\omega_{B}$, and their linewidths in free-space are $\Gamma_{A}$ and $\Gamma_{B}$, respectively.
Generally, the upper/lower states of each atom, $|A_{\pm}\rangle$, $|B_{\pm}\rangle$, may be degenerate. 
Further, in order to rest within the weak coupling regime and to ensure the distinction between the atomic species, the detuning
$\Delta_{AB}\equiv\omega_{A}-\omega_{B}$ is such that  $|\Delta_{AB}|\gg\langle W\rangle/\hbar,(\Gamma_{A}+\Gamma_{B})/2$,
with $W$ being the interaction Hamiltonian. In Schr\"odinger's representation, the time-propagator of the two-atom-EM field system reads, within a time interval  $[T_{0},T]$, $\mathbb{U}(T-T_{0})=\mathcal{T}\exp{}\left[-i\hbar^{-1}\int_{T_{0}}^{T}\textrm{d}t\left( H_{0}+W\right)\right]$.
%\begin{equation}
%\mathbb{U}(T-T_{0})=\mathcal{T}\exp{}\left[-i\hbar^{-1}\int_{T_{0}}^{T}\textrm{d}t\left( H_{0}+W\right)\right].\nonumber
%\end{equation}
In this expression $H_{0}$ is the sum of the free Hamiltonians of the internal atomic states and of the EM field, $H_{0}=\hbar\omega_{A}|A_{+}\rangle\langle A_{+}|+\hbar\omega_{B}|B_{+}\rangle\langle B_{+}|+\sum_{\mathbf{k},\mathbf{\epsilon}}\hbar\omega(a^{\dagger}_{\mathbf{k},\mathbf{\epsilon}}a_{\mathbf{k},\mathbf{\epsilon}}+1/2)$,
%\begin{equation}
%H_{EM}=\sum_{\mathbf{k},\mathbf{\epsilon}}\hbar\omega(a_{\mathbf{k},\mathbf{\epsilon}}a^{\dagger}_{\mathbf{k},\mathbf{\epsilon}}+1/2),\nonumber
%\end{equation}
where $\omega=ck$ is the photon frequency, and the operators $a^{\dagger}_{\mathbf{k},\mathbf{\epsilon}}$ and $a_{\mathbf{k},\mathbf{\epsilon}}$ are the creation and annihilation operators of photons with momentum $\hbar\mathbf{k}$ and polarization $\mathbf{\epsilon}$ respectively. Finally, the interaction Hamiltonian in the electric dipole approximation reads $W=W_{A}+W_{B}$, with $W_{A,B}=-\mathbf{d}_{A,B}\cdot\mathbf{E}(\mathbf{R}_{A,B})$. In this expression $\mathbf{d}_{A,B}$ are the electric dipole operators of each atom and  $\mathbf{E}(\mathbf{R}_{A,B})$ is the electric field operator evaluated at the position of each atom, which   can be written in the usual manner as a sum over normal modes,
\begin{eqnarray}\label{AQ}
\mathbf{E}(\mathbf{R}_{A,B})&=&\sum_{\mathbf{k}}\mathbf{E}^{(-)}_{\mathbf{k}}(\mathbf{R}_{A,B})+\mathbf{E}^{(+)}_{\mathbf{k}}(\mathbf{R}_{A,B})\nonumber\\
&=&i\sum_{\mathbf{k},\mathbf{\epsilon}}\sqrt{\frac{\hbar ck}{2\mathcal{V}\epsilon_{0}}}
[\mathbf{\epsilon}a_{\mathbf{k},\epsilon}e^{i\mathbf{k}\cdot\mathbf{R}_{A,B}}-\mathbf{\epsilon}^{*}a^{\dagger}_{\mathbf{k},\epsilon}e^{-i\mathbf{k}\cdot\mathbf{R}_{A,B}}],\nonumber
\end{eqnarray}
where $\mathcal{V}$ is a volume of quantisation and $\mathbf{E}^{(\mp)}_{\mathbf{k}}$ denote the annihilation and creation electric field operators of photons of momentum $\hbar\mathbf{k}$, respectively. Correspondingly, we define $W_{A,B;\mathbf{k}}^{(\mp)}=-\mathbf{d}_{A,B}\cdot\mathbf{E}_{\mathbf{k}}^{(\mp)}(\mathbf{R}_{A,B})$. 

Next, considering $W$ as a perturbation to the free Hamiltonians, the unperturbed time-propagator for free atom and free photon states is, within the time interval $[t',t]$, $\mathbb{U}_{0}(t-t')=\exp{[-i\hbar^{-1}H_{0}(t-t')]}$. In terms of $W$ and $\mathbb{U}_{0}$, $\mathbb{U}(T-T_{0})$ admits an expansion in powers of  $W$ which can be developed out of the time-ordered exponential equation,
\begin{equation}
\mathbb{U}(T-T_{0})=\mathbb{U}_{0}(T)\:\mathcal{T}\exp\int_{T_{0}}^{T}(-i/\hbar)\mathbb{U}_{0}^{\dagger}(t)\:W\:\mathbb{U}_{0}(t-T_{0})\textrm{d}t.\nonumber
\end{equation}
Denoting the term of order $W^{n}$ in the corresponding series by $\delta\mathbb{U}^{(n)}$, we may write $\mathbb{U}(T-T_{0})=\mathbb{U}_{0}(T-T_{0})+\sum_{n=1}^{\infty}\delta\mathbb{U}^{(n)}(T-T_{0})$.  We denote by $|\Psi_{0}\rangle$ the state of the system at the initial time $T_{0}$, that we take equal to 0 for simplicity. Consequently, the state of the system at the time of observation $T$ is $|\Psi(T)\rangle=\mathbb{U}(T)|\Psi_{0}\rangle$. Hereafter we will omit $T$, unless necessary, in the argument of expectation values.
%As an example, $\delta\mathbb{U}^{(2)}$ reads
%\begin{equation}
%\delta\mathbb{U}^{(2)}(T-T_{0})=-\hbar^{2}\int_{0}^{T}\textrm{d}t\int_{0}^{t}\textrm{d}t'\mathbb{U}_{0}(T-t)W\mathbb{U}_{0}(t-t')W\mathbb{U}_{0}(t'-T_{0}).
%\end{equation}

%Generically, the initial state of the system at time $T_{0}$ will be denoted by $|\Psi_{0}\rangle$. This being the case,Correspondingly, in regards to the computation of any physical observable at times $T\geq\pi/\Omega$, the asymptotic state of the system in the far past may be approximated by , and the coupling to the vacuum EM field proceeds adiabatically. The latter is implemented mathematically by making the substitutions $T_{0}\rightarrow-\infty$ and $W\rightarrow W\exp{\eta t}$, with $\eta\rightarrow0^{+}$, in the equation for $\mathbb{U}(T-T_{0})$.

\begin{figure}[h]
\includegraphics[height=10.7cm,width=8.7cm,clip]{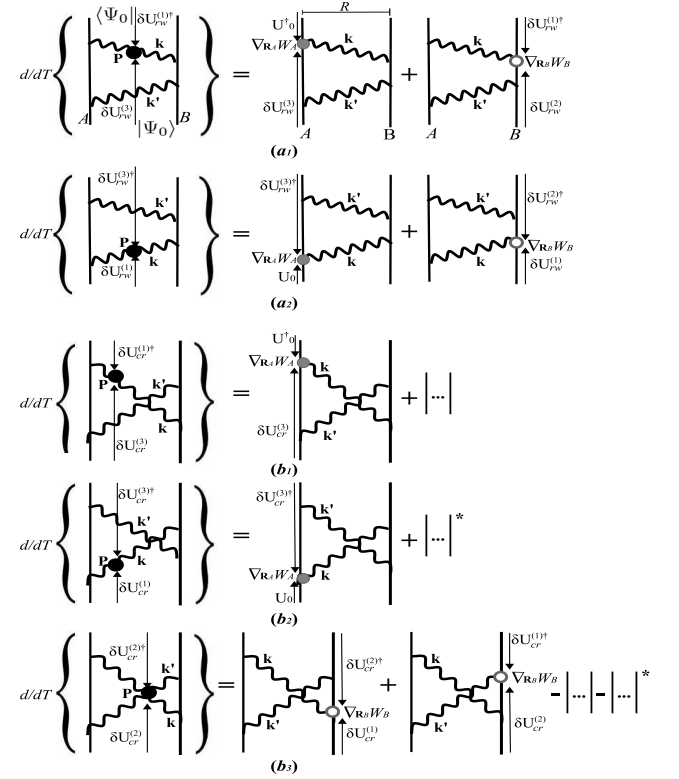}
\caption{Diagrammatic representation of Eqs.(\ref{EQ1p}) [$(a_{1})$ and $(a_{2})$] , (\ref{EQ2ap}) [$(b_{1})$ and $(b_{2})$], and (\ref{EQ2bp}) [$(b_{3})$].
Thick straight lines stand for time-propagators of atomic states, while wavy lines do for space-time-propagators of photons of momenta $\hbar\mathbf{k}$ and $\hbar\mathbf{k}'$. The atoms are separated by a distance $R$ along
the horizontal direction, whereas time runs along the vertical. Black circles stand for the insertion of an EM momentum operator, $\mathbf{P}$, grey circles  stand for $\nabla_{\mathbf{R}_{A}}W_{A}$, and white circles do for $\nabla_{\mathbf{R}_{B}}W_{B}$. In each diagram two time-propagators (depicted by vertical arrows) evolve the intial states, $|\psi_{0}\rangle$ and $\langle\psi_{0}|$, towards the observation time $T$, at which one of the aforementioned operators applies.}\label{figure1_AR}
\end{figure}

Straight application of the definition of the force operator on each atom yields for the total force on the system,
\begin{align}
\langle\mathbf{F}_{A}&+\mathbf{F}_{B}\rangle=\langle\dot{\mathbf{Q}}_{A}+\dot{\mathbf{Q}}_{B}\rangle=-i\hbar\partial_{T}\langle\Psi_{0}|\mathbb{U}^{\dagger}(T)(\mathbf{\nabla}_{\mathbf{R}_{A}}\nonumber\\
&+\mathbf{\nabla}_{\mathbf{R}_{B}})\mathbb{U}(T)|\Psi_{0}\rangle=-\langle \mathbf{\nabla}_{\mathbf{R}_{A}}W_{A}\rangle-\langle\mathbf{\nabla}_{\mathbf{R}_{B}}W_{B}\rangle, \label{Force}
\end{align}
where $\mathbf{Q}_{A,B}$ are the kinetic momentum operators of the centers of mass of each atom, and terms linear in $v_{A,B}/c$, with $v_{A,B}$ the velocities of the atoms, have been discarded \cite{MyPRAv}. The diagrammatical representation of $\langle W_{A,B}\rangle$ at $\mathcal{O}(W_{A}^{2}W_{B}^{2})$ was already given in Ref.\cite{Me2PRA}. %Next and for the sake of simplicity, let us restrict ourselves to the quasiresonant approximation $(qr)$, in which the dominant contribution to the forces comes from the diagram of Fig.\ref{figure1}. 
 It was argued there, and also in precedent articles \cite{McLone,Power1965,Power1995}, that only the two kinds of diagrams in Figs.\ref{figure1_AR}$(a_{1},a_{2})$ and \ref{figure1_AR}$(b_{1}-b_{3})$ contribute to the resonant forces after an adiabatic excitation. Thus, they are the only ones that may cause the non-vanishing of Eq.(\ref{Force}). On the other hand, in virtue of total momentum conservation, a non-zero total force must be compensated by an opposite variation of the momentum of the EM vacuum. In the following we show over the relevant diagrams that this is indeed the case. That is, being $\mathbf{P}=\sum_{\mathbf{k},\mathbf{\epsilon}}\hbar\mathbf{k}(a^{\dagger}_{\mathbf{k},\mathbf{\epsilon}}a_{\mathbf{k},\mathbf{\epsilon}}+1/2)$ the momentum operator of the EM field, we show that $\langle\dot{\mathbf{P}}\rangle=-\langle\mathbf{F}_{A}+\mathbf{F}_{B}\rangle $ 
%\begin{equation}\label{LAEQ}
%\langle\dot{\mathbf{P}}\rangle=-[\langle\mathbf{F}_{A}\rangle+\langle\mathbf{F}_{B}\rangle] 
%&=\langle\mathbf{\nabla}_{\mathbf{R}_{A}}W_{A}(T)\rangle+\langle\mathbf{\nabla}_{\mathbf{R}_{B}}W_{B}(T)\rangle\nonumber\\
%\end{equation}
at zero order in $v_{A,B}/c$  \footnote{In addition to the transverse momentum $\mathbf{P}$ there exists also a longitudinal momentum operator \cite{PRLMoMe} which, in the electric dipole approximation, equals the R\"ontgen momentum \cite{MyPRAv}. Consistently with the neglect of terms of order $v_{A,B}/c$ in the total force, it is discarded here.}. Hence, violation of the action-reaction principle implies a non-zero vacuum momentum and viceversa. Hereafter, we will refer to the quantities associated to the diagram of Figs.\ref{figure1_AR}($a_{1},a_{2}$) as \emph{rotating-wave} ($rw$), since they scale as $\sim\Delta_{AB}^{-1}$. Their contribution dominates under quasiresonant conditions, $|\Delta_{A,B}|\gg\omega_{A,B}$. Likewise, the quantities associated to the diagram of Figs.\ref{figure1_AR}($b_{1}-b_{3}$) will be referred to as \emph{counter-rotating} ($cr$), as they go like $\sim(\omega_{A}+\omega_{B})^{-1}$.

In the first place, the time derivative of the one-photon momentum associated to the rotating diagram reads
\begin{align}
\langle\dot{\mathbf{P}}\rangle^{1\gamma}_{rw}&=\sum_{\mathbf{k}}\partial_{T}\langle\Psi_{0}|\delta\mathbb{U}_{rw}^{\dagger(1)}(T,\mathbf{k})\hbar\mathbf{k}\delta\mathbb{U}_{rw}^{(3)}(T,\mathbf{k})|\Psi_{0}\rangle+\textrm{c.c.},\label{EQ1}
\end{align}
where the first term corresponds to the diagram on the l.h.s. of Fig.\ref{figure1_AR}$(a_{1})$, and its complex conjugate (c.c.) does to the diagram on the l.h.s. of Fig.\ref{figure1_AR}$(a_{2})$. The expressions of the time-propagators appearing in Eq.(\ref{EQ1}) and thereafter are compiled in the Appendix A.
Performing the time derivative over the propagators we arrive at
\begin{align}
\langle\dot{\mathbf{P}}\rangle^{1\gamma}_{rw}&=\sum_{\mathbf{k}}\Bigl[\langle\Psi_{0}|\mathbb{U}^{\dagger}_{0}(T)i\mathbf{k}W_{A,\mathbf{k}}^{(-)}\delta\mathbb{U}_{rw}^{(3)}(T,\mathbf{k})|\Psi_{0}\rangle\nonumber\\
&+\langle\Psi_{0}|\delta\mathbb{U}^{\dagger(1)}_{rw}(T,\mathbf{k})(-i\mathbf{k})W_{B,\mathbf{k}}^{(+)}\delta\mathbb{U}_{rw}^{(2)}(T)|\Psi_{0}\rangle\Bigr]
+\textrm{c.c.}\nonumber\\
&=\langle\mathbf{\nabla}_{\mathbf{R}_{A}}W_{A}\rangle_{rw}+\langle\mathbf{\nabla}_{\mathbf{R}_{B}}W_{B}\rangle_{rw}.\label{EQ1p}
\end{align}
The first two terms on the r.h.s. of Eq.(\ref{EQ1p}) correspond to the two diagrams on the r.h.s. of Fig.\ref{figure1_AR}$(a_{1})$, respectively. Likewise, their complex conjugates correspond to the  diagrams on the r.h.s. of Fig.\ref{figure1_AR}$(a_{2})$. 

As for the time derivative of the vacuum momentum associated to the counter-rotating diagram, it contains both one-photon and two-photon components. The derivative of the one-photon momentum is
\begin{align}
\langle\dot{\mathbf{P}}\rangle^{1\gamma}_{cr}&=\sum_{\mathbf{k}}\partial_{T}\langle\Psi_{0}|\delta\mathbb{U}_{cr}^{\dagger(1)}(T,\mathbf{k})\hbar\mathbf{k}\delta\mathbb{U}_{cr}^{(3)}(T,\mathbf{k})|\Psi_{0}\rangle+\textrm{c.c.}\label{EQ2a}
\end{align}
The first term on the r.h.s. of Eq.(\ref{EQ2a}) corresponds to the diagram on the l.h.s. of Fig.\ref{figure1_AR}$(b_{1})$, while its complex conjugate does to the diagram on the l.h.s. of Fig.\ref{figure1_AR}$(b_{2})$. Derivation with respect to $T$ yields
\begin{align}
\langle\dot{\mathbf{P}}\rangle^{1\gamma}_{cr}&=\sum_{\mathbf{k}}\langle\Psi_{0}|\mathbb{U}^{\dagger}_{0}(T)i\mathbf{k}W_{A,\mathbf{k}}^{(-)}\delta\mathbb{U}_{cr}^{(3)}(T,\mathbf{k})|\Psi_{0}\rangle\nonumber\\
&+\mathcal{J}(T)+\textrm{c.c.}=\langle\mathbf{\nabla}_{\mathbf{R}_{A}}W_{A}\rangle_{cr}+2\textrm{Re}\mathcal{J}(T),\label{EQ2ap}
\end{align}
where $\mathcal{J}(T)$ is a function of $T$ non-relevant to us. The first term on the r.h.s. of Eq.(\ref{EQ2ap}) corresponds to the diagram on the r.h.s. of Fig.\ref{figure1_AR}$(b_{1})$, while its complex conjugate corresponds to the  diagram on the r.h.s. of Fig.\ref{figure1_AR}$(b_{2})$. The function $\mathcal{J}(T)$ is depicted by ellipsis there. Regarding the variation of the two-photon momentum,
\begin{align}
\langle\dot{\mathbf{P}}\rangle^{2\gamma}_{cr}&=\sum_{\mathbf{k},\mathbf{k}'}\partial_{T}\langle\Psi_{0}|\delta\mathbb{U}_{cr}^{\dagger(2)}(T,\mathbf{k}',\mathbf{k})\hbar(\mathbf{k}+\mathbf{k}')\nonumber\\
&\times\delta\mathbb{U}_{cr}^{(2)}(T,\mathbf{k},\mathbf{k}')|\Psi_{0}\rangle,\label{EQ2b}
\end{align}
it is represented by the diagram on the l.h.s. of Fig.\ref{figure1_AR}($b_{3}$). 
Performing the time derivative of Eq.(\ref{EQ2b}) we arrive at
\begin{align}
\langle\dot{\mathbf{P}}\rangle^{2\gamma}_{cr}&=\sum_{\mathbf{k}}\langle\Psi_{0}|\delta\mathbb{U}_{cr}^{\dagger(2)}(T,\mathbf{k}',\mathbf{k})(-i\mathbf{k})W_{B,\mathbf{k}}^{(+)}\delta\mathbb{U}_{cr}^{(1)}(T,\mathbf{k})|\Psi_{0}\rangle\nonumber\\
&-\mathcal{J}(T)+\textrm{c.c.}=\langle\mathbf{\nabla}_{\mathbf{R}_{B}}W_{B}\rangle_{cr}-2\textrm{Re}\mathcal{J}(T).\label{EQ2bp}
\end{align}
The first term on the r.h.s. of Eq.(\ref{EQ2bp}) is depicted by the first diagram on the r.h.s. of Fig.\ref{figure1_AR}$(b_{3})$, while its complex conjugate corresponds to the second diagram there. The term $-2\textrm{Re}\mathcal{J}(T)$ cancels the last term of Eq.(\ref{EQ2ap}). Interestingly, the counter-rotating one-photon momentum compensates the gain of momentum by atom $A$, while the two-photon momentum compensates the momentum of atom $B$. Eq.(\ref{EQ2bp}), together with (\ref{EQ2ap}) and  (\ref{EQ1p}), complete the proof of the equality $\langle\dot{\mathbf{P}}\rangle=-\langle\mathbf{F}_{A}+\mathbf{F}_{B}\rangle$.

In a strict manner, in order to capture the complete dynamics of the forces it would be necessary to include in $\mathbb{U}_{0}$ the dynamical excitation of atom $A$. Nonetheless,  it was shown in Ref.\cite{Me2PRA} that for the usual case of an adiabatic excitation driven by a $\pi$-pulse of frequency $\Omega$, with $|\Delta_{AB}|\gg\Omega\gg\Gamma_{A}$, the interaction becomes quasistationary at times of observation $T\geq\pi/\Omega$. 
Mathematically, same quasistationary result is obtained if $|\Psi_{0}\rangle$ is considered as an asymptotic state in the far past, $|\Psi_{0}\rangle=|A_{+}\rangle\otimes|B_{-}\rangle\otimes|0_{\gamma}\rangle$, $|0_{\gamma}\rangle$ being the EM vacuum state, and the substitutions 
 $T_{0}\rightarrow-\infty$, $W\rightarrow W\exp{\eta t}$, with $\eta\rightarrow0^{+}$, are made in the equation for $\mathbb{U}(T-T_{0})$. In so doing we find for the total force \cite{Me2PRA}
\begin{align} 
-\langle\dot{\mathbf{P}}\rangle&\simeq\langle\mathbf{F}_{A}+\mathbf{F}_{B}\rangle_{0}e^{-\Gamma_{A}T}=\mathcal{U}^{ijpq}e^{-\Gamma_{A}T}\nonumber\\
&\times\nabla_{\mathbf{R}}[\textrm{Im}G_{ij}^{(0)}(\mathbf{R},\omega_{A})\textrm{Im}G_{pq}^{(0)}(\mathbf{R},\omega_{A})]/k_{A}^{3}.\label{Forces}
\end{align}
In this equation $\mathbb{G}^{(0)}(\mathbf{R},\omega)$ is the dyadic Green's function of the electric field induced at $\mathbf{R}=\langle\mathbf{R}_{A}-\mathbf{R}_{B}\rangle$ by an electric dipole of frequency $\omega=ck$, 
\begin{equation}
\mathbb{G}^{(0)}(\mathbf{R},\omega)=\frac{k\:e^{ikR}}{-4\pi}[\alpha/kR+i\beta/(kR)^{2}-\beta/(kR)^{3}],
\end{equation}
where the tensors $\alpha$ and $\beta$ read $\alpha=\mathbb{I}-\mathbf{R}\mathbf{R}/R^{2}$,  $\beta=\mathbb{I}-3\mathbf{R}\mathbf{R}/R^{2}$, and $\mathcal{U}^{ijpq}=4\omega_{B}k^{7}_{A}\mu_{A}^{i}\mu_{A}^{q}\mu_{B}^{j}\mu_{B}^{p}/[\epsilon_{0}^{2}\hbar(\omega_{A}^{2}-\omega_{B}^{2})]$, with 
$\mu_{A}^{i}=\langle A_{-}|d^{i}_{A}|A_{+}\rangle$, $\mu_{B}^{j}=\langle B_{-}|d^{j}_{B}|B_{+}\rangle$. 

\begin{figure}[h]
\includegraphics[height=3.9cm,width=8.6cm,clip]{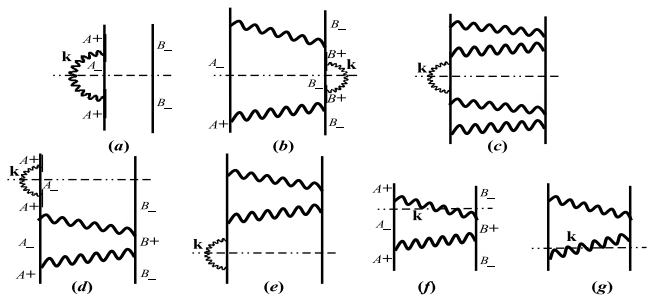}
\caption{Diagrammatic representation of $\mathcal{P}_{1\gamma}(\mathbf{k})$ in the quasiresonant approximation. The horizontal dashed lines intersect the 'will-be' emitted photon at the time of observation. Diagrams $(d)$, $(e)$ result from the interference of $(a)$ and $(c)$, whereas $(f)$, $(g)$ do so from $(a)$ and $(b)$.}\label{figure2_AR}
\end{figure}

We show next that the non-vanishing of Eq.(\ref{Forces}) results from the asymmetric interference of virtual photons along the interatomic direction, which ultimately implies the directionality of spontaneous emission. It is apparent from the expressions on the r.h.s. of Eqs.(\ref{EQ1}),  (\ref{EQ2a}) and (\ref{EQ2b}) that the photonic momenta may be factored out of the time derivatives there. The remaining quantities are the partial emission rates, of one and two photons, which are not invariant under the transformation $\mathbf{k}^{(')}\rightarrow-\mathbf{k}^{(')}$. Hence, we can write 
\begin{equation}\label{la13}
\langle\dot{\mathbf{P}}\rangle=\sum_{\mathbf{k},\mathbf{k}'}\hbar\mathbf{k}\dot{\mathcal{P}}_{1\gamma}(\mathbf{k})+\hbar(\mathbf{k}+\mathbf{k}')\dot{\mathcal{P}}_{2\gamma}(\mathbf{k},\mathbf{k}'),
\end{equation}
with $\mathcal{P}_{1\gamma}(\mathbf{k})=|\langle A_{-},B_{-},\gamma_{k}|\mathbb{U}(T)|\Psi_{0}\rangle|^{2}$ 
and $\mathcal{P}_{2\gamma}(\mathbf{k},\mathbf{k}')=|\langle A_{-},B_{-},\gamma_{k},\gamma_{k'}|\mathbb{U}(T)|\Psi_{0}\rangle|^{2}$ 
 being the probabilities of emission of one and two photons of momenta $\hbar\mathbf{k}$ and $\hbar(\mathbf{k}+\mathbf{k}')$ respectively. In the following and for simplicity we restrict ourselves to the quasiresonant approximation. Whereas  the photons emitted from atom $A$ in free-space do not distinguish between left and right along any direction [diagram of Fig.\ref{figure2_AR}$(a)$],  in the presence of atom $B$ there is a probability for them to be absorbed by $B$ and to be later scattered in any direction [Fig.\ref{figure2_AR}$(b)$]. It may also happen that the photons be reabsorbed by $A$ and ultimately rescattered [Fig.\ref{figure2_AR}$(c)$]. In addition, there are interference processes [Figs.\ref{figure2_AR}$(d-g)$]. The probabilities of scattering off $B$ and rescattering off $A$ are however of order $\Gamma_{A,B}^{2}/\Delta_{AB}^{2}$, and thus negligible in comparison to those of the interference terms which are $\mathcal{O}(\Gamma_{A}/\Delta_{AB})$. As shown by Berman in Ref.\cite{BermanOptTh}, the optical theorem demands that the total  probability of the interference terms cancel out to guarantee probability conservation. That is, the interference of the photons emitted from $A$ in free-space with the  photons rescattered off $A$ in any direction [\ref{figure2_AR}$(d,e)$] must be compensated by the interference of the photons emitted from $A$ with the  photons scattered off $B$ along the interatomic direction $\mathbf{R}$ [\ref{figure2_AR}$(f,g)$] --see Appendix B. The crucial point is that for this to be the case, the interference along $\mathbf{R}$ is necessarily asymmetric, and so is the distribution of photonic momentum which enters Eq.(\ref{la13}). In particular, the contribution of diagrams \ref{figure2_AR}$(f)$ and \ref{figure2_AR}$(g)$ to $\mathcal{P}_{1\gamma}(\mathbf{k})$ is, for asymptotic times and per unit solid angle  --see Appendix B and Ref.\cite{BermanRecoil},
\begin{equation}
\textrm{Re}\:\frac{k_{A}^{5}e^{ik_{A}R\cos{\theta}}}{(2\pi\epsilon_{0}\hbar)^{2}\Gamma_{A}\Delta_{AB}}\mu_{A}^{i}\mu_{B}^{j}(\delta_{ij}-\hat{k}_{i}\hat{k}_{j})\mu_{B}^{p}\mu_{A}^{q}G^{(0)}_{pq}(k_{A}R),\label{laqfalta}
\end{equation} 
where $\hat{\mathbf{k}}$ is a unitary vector along the emission direction and $\theta$ is the angle between $\hat{\mathbf{k}}$ and $\mathbf{R}$. In the far field, discarding the dependence on dipole orientations, this probability goes like $\sim\cos{[k_{A}R(\cos{\theta}+1)]}/k_{A}R$,  which gives rise to a maximal differential probability between forward and backward emission at $R\approx1.2/k_{A}$. %Before the deexcitation takes place, conservation of total momentum implies a non-vanishing total force on the two-atom system.
%As for the asymmetric interference beyond the quasiresonant approximation, it is represented by Fig.SM1 in the Supplemental Material.
 
Ultimately, when an actual photon is emitted, the excess of vacuum momentum along $\mathbf{R}$ results in the directionality of spontaneous emission. Thus, action-reaction violation implies directional emission. %It is worth noting that the lack of mirror symmetry is not originated by the different nature of the atoms, but by their asymmetric excitation. 
 Hence, Eq.(\ref{Forces}) yields the average momentum of the photons emitted at $T\rightarrow\infty$,  $\langle\mathbf{P}\rangle_{\infty}=-\langle\mathbf{F}_{A}+\mathbf{F}_{B}\rangle_{0}/\Gamma_{A}$.% where for weak coupling $\Gamma_{A}$ equals the free-space emission rate.
Recent publications have drawn attention to the directional excitation of guided modes using the directional emission of excited atoms. Such is the case of the experimental and theoretical works of Mitsch \emph{et al.} \cite{Mitsch} and Scheel \emph{et al.} \cite{Scheel}. In their setup directional emission depends on the polarization of the emitted light, and the ground state system is a nanofiber. Although similar effects to the ones described in this paper are likely to occur there, the theoretical study of action-reaction violation is problematic with regard to the quantum description of the coupling between the EM field and the macroscopic fiber. 

\begin{figure}[h]
\includegraphics[height=3.6cm,width=8.4cm,clip]{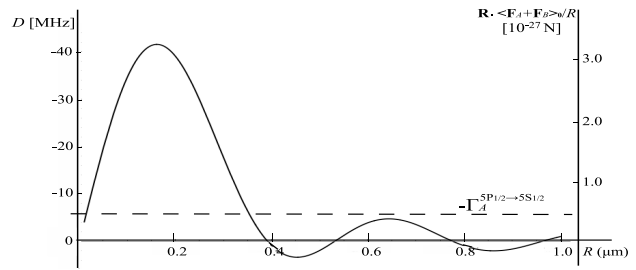}
\caption{Graphical representation, as a function of the interatomic distance $R$, of the directionality of spontaneous emission, $D$, and of the total force on a system composed by an atom of $^{87}$Rb in the state 5P$_{1/2}$ and a ground state $^{40}$K atom. }\label{figure3_AR}
\end{figure}

In the following, we compute the total force of Eq.(\ref{Forces}) on a system composed by an atom of $^{87}$Rb in the state 5P$_{1/2}$ and a $^{40}$K atom in its ground state. In principle, atomic forces can be measured through the displacement of each atom within their optical traps. %\cite{ReviewdeBrowaye}.% 
Alternatively, directional emission can be observed spectroscopically. It can be quantified through the average photon momentum emitted along $\mathbf{R}$,  $D=\mathbf{R}\cdot\langle\mathbf{P}\rangle_{\infty}c/Rh$ \cite{Scheel}. In Fig.\ref{figure3_AR} we plot the values of $D$ together with the total force on the system as a function of $R$. Maximum values are obtained for $R\simeq1.28/k_{A}$, where $D$ is $\sim10$ times greater than $\Gamma_{A}$, $\sim10^{6}$ greater than the vdW potentials, but yet $10^{7}$ times less than $\omega_{A}$.
%As anticipated, the addition of Eqs.(\ref{EQ2ap}) and (\ref{EQ2bp}) results in 
%\begin{equation}
%\partial_{T}[\langle\mathbf{P}^{EM}\rangle^{1\gamma}_{cr}+\langle\mathbf{P}^{EM}\rangle^{2\gamma}_{cr}]=-[\langle\mathbf{F}_{A}(T)\rangle+\langle\mathbf{F}_{B}(T)\rangle]_{cr}.
%\end{equation}

In this article we have shown that the vdW forces acting on a two-atom system, with one atom excited, are compensated by the transfer of momentum to the  EM vacuum, in accordance with conservation of total momentum. The non-vanishing of the vacuum momentum, and hence of the net force, is a manifestation of the optical theorem on the scattering of virtual photons. Ultimately, the vacuum momentum is released through directional spontaneous emission. The measure of directional emission is proposed as a means to quantify the violation of the action-reaction principle in the system. In contrast to the vacuum momentum found in Ref.\cite{PRLMoMe}, where the asymmetry under $\mathbf{k}\rightarrow-\mathbf{k}$ is caused by the presence of a chiral molecule and an external magnetic field, here it is a consequence of the optical theorem within an asymmetrically excited system. Beside our simple QED calculation, the optical theorem manifests itself in any radiation-mediated interaction. Therefore, directional emission and hence violation of  action-reaction is to be found in any asymmetrically excited system whose deexcitation proceeds through radiative emission of any kind.% In particular, it would be worth studying the kinematical effects of these phenomena on networks of long-life metastable topological defects \cite{Kibble-MeJPA-ArttuMonopoles,StringInt1-StringInt2}.

\acknowledgments
We thank P. Berman and M.-P. Gorza for useful discussions on this problem. 
Financial support from ANR-10-IDEX-0001-02-PSL  is gratefully acknowledged.
\appendix
\begin{widetext}
\section{Appendix A: Time-propagators in rotating  and counter-rotating processes}\label{AppA}

We give below the explicit equations for the time-propagator components of order $n$, $\delta\mathbb{U}_{rw}^{(n)}$ and  $\delta\mathbb{U}_{cr}^{(n)}$, which enter Eqs.(2-7) for the time derivative of the vacuum momentum, and appear on the \emph{rotating} and \emph{counter-rotating} diagrams of Fig.1.

As for Eq.(2) and Fig.1$(a_{1})$, $\delta\mathbb{U}_{rw}^{\dagger(1)}(T,\mathbf{k})$ and $\delta\mathbb{U}_{rw}^{(3)}(T,\mathbf{k})$ read, respectively,
\begin{eqnarray}
\delta\mathbb{U}_{rw}^{\dagger(1)}(T,\mathbf{k})&=&(i/\hbar)\int_{0}^{T}\textrm{d}t\:\mathbb{U}^{\dagger}_{0}(t)W_{A,\mathbf{k}}^{(-)}\mathbb{U}^{\dagger}_{0}(T-t),\nonumber\\
\delta\mathbb{U}_{rw}^{(3)}(T,\mathbf{k})&=&(-i/\hbar)^{3}\sum_{\mathbf{k}'}\int_{0}^{T}\textrm{d}t\int_{0}^{t}\textrm{d}t'\int_{0}^{t'}\textrm{d}t''
\mathbb{U}_{0}(T-t)W_{B,\mathbf{k}}^{(+)}\mathbb{U}_{0}(t-t')W_{B,\mathbf{k}'}^{(-)}\mathbb{U}_{0}(t'-t'')W_{A,\mathbf{k}'}^{(+)}\mathbb{U}_{0}(t'').\nonumber
\end{eqnarray}
The time derivative of the above quantities within Eq.(2) gives rise to Eq.(3), where the time-propagator $\delta\mathbb{U}_{rw}^{(2)}(T)$ is
\begin{equation}
\delta\mathbb{U}_{rw}^{(2)}(T)=(-i/\hbar)^{2}\sum_{\mathbf{k}'}\int_{0}^{T}\textrm{d}t\int_{0}^{t}\textrm{d}t'
\mathbb{U}_{0}(T-t) W_{B,\mathbf{k}'}^{(-)}\mathbb{U}_{0}(t-t')W_{A,\mathbf{k}'}^{(+)}\mathbb{U}_{0}(t').\nonumber
\end{equation}

As for Eq.(4) and Fig.1$(b_{1})$, $\delta\mathbb{U}_{cr}^{\dagger(1)}(T,\mathbf{k})=\delta\mathbb{U}_{rw}^{\dagger(1)}(T,\mathbf{k})$, and $\delta\mathbb{U}_{cr}^{(3)}(T,\mathbf{k})$ is
\begin{equation}
\delta\mathbb{U}_{cr}^{(3)}(T,\mathbf{k})=(-i/\hbar)^{3}\sum_{\mathbf{k}'}\int_{0}^{T}\textrm{d}t\int_{0}^{t}\textrm{d}t'\int_{0}^{t'}\textrm{d}t''
\mathbb{U}_{0}(T-t) W_{B,\mathbf{k}'}^{(-)}\mathbb{U}_{0}(t-t')W_{B,\mathbf{k}}^{(+)}\mathbb{U}_{0}(t'-t'')W_{A,\mathbf{k}'}^{(+)}\mathbb{U}_{0}(t'').\nonumber
\end{equation}

Finally, the  time-propagator $\delta\mathbb{U}_{cr}^{(2)}(T,\mathbf{k},\mathbf{k}')$ entering Eqs.(6) and Fig.1$(b_{3})$  reads
\begin{equation}
\delta\mathbb{U}_{cr}^{(2)}(T,\mathbf{k},\mathbf{k}')=(-i/\hbar)^{2}\int_{0}^{T}\textrm{d}t\int_{0}^{t}\textrm{d}t'
\mathbb{U}_{0}(T-t) W_{B,\mathbf{k}}^{(+)}\mathbb{U}_{0}(t-t')W_{A,\mathbf{k}'}^{(+)}\mathbb{U}_{0}(t').\nonumber
\end{equation}
Note that the propagator  $\delta\mathbb{U}_{cr}^{\dagger(2)}(T,\mathbf{k}',\mathbf{k})$ in Eqs.(6,7)  is the hermitian conjugate of the above expression upon which the exchange $\mathbf{k}\leftrightarrow\mathbf{k}'$ has been performed.

As for the function $\mathcal{J}(T)$, which enters Eqs.(5) and (7) with opposite signs, it results from the time derivatives of $\delta\mathbb{U}_{cr}^{(3)}(T,\mathbf{k})$ within Eq.(4) and of $\delta\mathbb{U}_{cr}^{\dagger(2)}(T,\mathbf{k}',\mathbf{k})$ within Eq.(6), respectively. It reads 
\begin{equation}
\mathcal{J}(T)=-\sum_{\mathbf{k},\mathbf{k}'}\langle\Psi_{0}|\delta\mathbb{U}_{rw}^{\dagger(1)}(T,\mathbf{k})i\mathbf{k}W_{B,\mathbf{k}'}^{(-)}\delta\mathbb{U}_{cr}^{(2)}(T,\mathbf{k},\mathbf{k}')|\Psi_{0}\rangle\nonumber.
\end{equation}

\section{Appendix B: Optical theorem in the quasiresonant approximation}\label{AppB}
It has been argued that satisfaction of the optical theorem guaranties the conservation of total emission probability. In the quasiresonant approximation, the leading order processes which contribute to the one-photon emission probability, $\mathcal{P}_{1\gamma}$, are represented diagrammatically in Fig.2. We show next that, at leading order in $\Gamma_{A,B}/\Delta_{AB}$, $\mathcal{P}_{1\gamma}=1$. This result was firstly obtained by Berman in Ref.\cite{BermanOptTh} using an equivalent formalism.

In the first place, the probability of emission in free-space reads, from diagram $2(a)$,
\begin{eqnarray}
\mathcal{P}_{1\gamma}^{(a)}&=&\sum_{\mathbf{k}}\mathcal{P}_{1\gamma}^{(a)}(\mathbf{k})=\sum_{\mathbf{k}}\langle\Psi_{0}|\delta\mathbb{U}_{rw}^{\dagger(1)}(T,\mathbf{k})\delta\mathbb{U}_{rw}^{(1)}(T,\mathbf{k})|\Psi_{0}\rangle\nonumber\\
&=&\mu_{A}^{i}\mu_{A}^{j}\hbar^{-2}\sum_{\mathbf{k}}\left|\int_{0}^{T}\textrm{d}t\:e^{-i\omega(T-t)}e^{-it\omega_{A}}e^{-t\Gamma_{A}/2}\right|^{2}
\langle0_{\gamma}|E^{(-)}_{\mathbf{k},i}(\mathbf{R}_{A})E^{(+)}_{\mathbf{k},j}(\mathbf{R}_{A})|0_{\gamma}\rangle.\nonumber
\end{eqnarray}
Perfoming the time integration, passing the sum over $\mathbf{k}$ to a continuous integral and performing the resultant integration over $\mathbf{k}$, we obtain for asymptotic times $\Gamma_{A}T\gg1$,
\begin{equation}
\mathcal{P}_{1\gamma}^{(a)}\simeq\frac{-2k^{2}_{A}}{\epsilon_{0}\hbar\Gamma_{A}}\mu_{A}^{i}\mu_{A}^{j}\textrm{Im}G^{(0)}_{ij}(\mathbf{r},\omega_{A})=1, \textrm{ for }\mathbf{r}\rightarrow\mathbf{0},\nonumber
\end{equation}
where in the last equality we have identified $\Gamma_{A}$, by consistency, with the free-space emission rate. Therefore, probability conservation demands the cancellation at the lowest order of the total emission coming from the rest of diagrams in Fig.2.

As for the emission probability of diagrams $2(d)$ and $2(e)$, which result from the interference of $2(a)$ and $2(c)$, we find for asymptotic times,
\begin{eqnarray}
\mathcal{P}_{1\gamma}^{(d,e)}&=&\sum_{\mathbf{k}}\mathcal{P}_{1\gamma}^{(c,d)}(\mathbf{k})=2\textrm{Re}\:\mu_{A}^{i}\mu_{A}^{j}\mu_{A}^{p}\mu_{B}^{q}\mu_{B}^{r}\mu_{A}^{s}\hbar^{-6}\sum_{\mathbf{k},\mathbf{k}',\mathbf{k}''}\int_{0}^{T}\textrm{d}t\int_{0}^{T}\textrm{d}t'\int_{0}^{t'}\textrm{d}t''\int_{0}^{t''}\textrm{d}t'''
\int_{0}^{t'''}\textrm{d}t^{iv}\int_{0}^{t^{iv}}\textrm{d}t^{v}e^{i\omega_{A}t}e^{-t\Gamma_{A}/2}\nonumber\\
&\times&e^{-i(t-t')\omega}e^{-i(t'-t'')\omega_{A}}e^{-(t'-t'')\Gamma_{A}/2}e^{-i(t''-t''')\omega'}e^{-i(t'''-t^{iv})\omega_{B}}e^{-(t'''-t^{iv})\Gamma_{B}/2}e^{-i(t^{iv}-t^{v})\omega''}e^{-it^{v}\omega_{A}}e^{-t^{v}\Gamma_{A}/2}\nonumber\\
&\times&\langle0_{\gamma}|E^{(-)}_{\mathbf{k},i}(\mathbf{R}_{A})E^{(+)}_{\mathbf{k},j}(\mathbf{R}_{A})|0_{\gamma}\rangle\langle0_{\gamma}|E^{(-)}_{\mathbf{k}',p}(\mathbf{R}_{A})E^{(+)}_{\mathbf{k}',q}(\mathbf{R}_{B})|0_{\gamma}\rangle\langle0_{\gamma}|E^{(-)}_{\mathbf{k}'',r}(\mathbf{R}_{B})E^{(+)}_{\mathbf{k}'',s}(\mathbf{R}_{A})|0_{\gamma}\rangle\nonumber\\
&\simeq&\frac{4 k_{A}^{4}}{\epsilon^{2}_{0}\hbar^{2}\Gamma_{A}\Delta_{AB}}\mu_{A}^{p}\textrm{Im}G^{(0)}_{pq}(\mathbf{R},\omega_{A})\mu_{B}^{q}\mu_{B}^{r}\textrm{Re}G^{(0)}_{rs}(\mathbf{R},\omega_{A})\mu_{A}^{s}.\nonumber
\end{eqnarray}

On the other hand, the partial emission probability of one photon of momentum $\hbar k_{A}$  along a generic direction $\hat{\mathbf{k}}$, given by Eq.(\ref{laqfalta}), derives from the partial integration of the emission probability of diagrams $2(f)$ and $2(g)$, which are the result of the interference of the photons emitted in diagrams  $2(a)$ and $2(b)$ along $\mathbf{R}$. The expression for the total emission probability of diagrams $2(f)$ and $2(g)$ is
\begin{eqnarray}
\mathcal{P}_{1\gamma}^{(f,g)}&=&\sum_{\mathbf{k}}\mathcal{P}_{1\gamma}^{(f,g)}(\mathbf{k})=2\textrm{Re}\sum_{\mathbf{k}}\langle\Psi_{0}|\delta\mathbb{U}_{rw}^{\dagger(1)}(T,\mathbf{k})\delta\mathbb{U}_{rw}^{(3)}(T,\mathbf{k})|\Psi_{0}\rangle\nonumber\\
&=&-2\textrm{Re}\:\mu_{A}^{p}\mu_{B}^{q}\mu_{B}^{r}\mu_{A}^{s}\hbar^{-4}\sum_{\mathbf{k},\mathbf{k}'}\int_{0}^{T}\textrm{d}t\int_{0}^{T}\textrm{d}t'\int_{0}^{t'}\textrm{d}t''\int_{0}^{t''}\textrm{d}t'''\:e^{i\omega_{A}t}e^{-t\Gamma_{A}/2}e^{-i(t-t')\omega}e^{-i(t'-t'')\omega_{B}}e^{-(t'-t'')\Gamma_{B}/2}\nonumber\\
&\times&e^{-i(t''-t''')\omega'}e^{-it'''\omega_{A}}e^{-t'''\Gamma_{A}/2}\langle0_{\gamma}|E^{(-)}_{\mathbf{k},p}(\mathbf{R}_{A})E^{(+)}_{\mathbf{k},q}(\mathbf{R}_{B})|0_{\gamma}\rangle\langle0_{\gamma}|E^{(-)}_{\mathbf{k}',r}(\mathbf{R}_{B})E^{(+)}_{\mathbf{k}',s}(\mathbf{R}_{A})|0_{\gamma}\rangle\nonumber\\
&\simeq&\frac{-4 k_{A}^{4}}{\epsilon^{2}_{0}\hbar^{2}\Gamma_{A}\Delta_{AB}}\mu_{A}^{p}\textrm{Im}G^{(0)}_{pq}(\mathbf{R},\omega_{A})\mu_{B}^{q}\mu_{B}^{r}\textrm{Re}G^{(0)}_{rs}(\mathbf{R},\omega_{A})\mu_{A}^{s},\nonumber
\end{eqnarray}
from which the relation $\mathcal{P}_{1\gamma}^{(d,e)}+\mathcal{P}_{1\gamma}^{(f,g)}=0$ holds as anticipated.

Finally, we show for the sake of completeness that the total probabilities of scattering off atom $B$ and rescattering off atom $A$ in any direction, depicted by the diagrams $2(b)$ and $2(c)$ respectively, are of an order $\Gamma_{A,B}/\Delta_{AB}$ smaller than the previous ones. As for scattering off $B$, its probability is
\begin{eqnarray}
\mathcal{P}_{1\gamma}^{(b)}&=&\sum_{\mathbf{k}}\mathcal{P}_{1\gamma}^{(b)}(\mathbf{k})=\sum_{\mathbf{k}}\langle\Psi_{0}|\delta\mathbb{U}_{rw}^{\dagger(3)}(T,\mathbf{k})\delta\mathbb{U}_{rw}^{(3)}(T,\mathbf{k})|\Psi_{0}\rangle\nonumber\\
&=&\mu_{B}^{i}\mu_{B}^{j}\hbar^{-6}\sum_{\mathbf{k}}\Bigl|\sum_{\mathbf{k}'}\int_{0}^{T}\textrm{d}t\int_{0}^{t}\textrm{d}t'\int_{0}^{t'}\textrm{d}t''e^{-i(T-t)\omega}e^{-i(t-t')\omega_{B}}e^{-(t-t')\Gamma_{B}/2}e^{-i(t'-t'')\omega'}e^{-it''\omega_{A}}e^{-t''\Gamma_{A}/2}\nonumber\\
&\times&\mu_{B}^{p}\mu_{A}^{q}\langle0_{\gamma}|E^{(-)}_{\mathbf{k}',p}(\mathbf{R}_{B})E^{(+)}_{\mathbf{k}',q}(\mathbf{R}_{A})|0_{\gamma}\rangle\Bigr|^{2}\langle0_{\gamma}|E^{(-)}_{\mathbf{k},i}(\mathbf{R}_{B})E^{(+)}_{\mathbf{k},j}(\mathbf{R}_{B})|0_{\gamma}\rangle\nonumber\\
&\simeq&\frac{\mu_{A}^{p}\mu_{B}^{q}\mu_{B}^{r}\mu_{A}^{s}}{\epsilon^{2}_{0}\hbar^{2}\Delta^{2}_{AB}}\left[\frac{|\mu_{B}|^{2}}{|\mu_{A}|^{2}}k_{A}^{4}G^{(0)}_{pq}(\mathbf{R},\omega_{A})G^{*(0)}_{rs}(\mathbf{R},\omega_{A})+k_{B}^{4}G^{(0)}_{pq}(\mathbf{R},\omega_{B})G^{*(0)}_{rs}(\mathbf{R},\omega_{B})\right],\nonumber
\end{eqnarray}
and for rescattering off $A$ we find
\begin{eqnarray}
\mathcal{P}_{1\gamma}^{(c)}&=&\sum_{\mathbf{k}}\mathcal{P}_{1\gamma}^{(c)}(\mathbf{k})=\mu_{A}^{i}\mu_{A}^{j}\hbar^{-10}\sum_{\mathbf{k}}\Bigl|\sum_{\mathbf{k}',\mathbf{k}''} \int_{0}^{T}\textrm{d}t'\int_{0}^{t'}\textrm{d}t''\int_{0}^{t''}\textrm{d}t'''
\int_{0}^{t'''}\textrm{d}t^{iv}\int_{0}^{t^{iv}}\textrm{d}t^{v}\nonumber\\
&\times&e^{-i(T-t')\omega}e^{-i(t'-t'')\omega_{A}}e^{-(t'-t'')\Gamma_{A}/2}e^{-i(t''-t''')\omega'}e^{-i(t'''-t^{iv})\omega_{B}}e^{-(t'''-t^{iv})\Gamma_{B}/2}e^{-i(t^{iv}-t^{v})\omega''}e^{-it^{v}\omega_{A}}e^{-t^{v}\Gamma_{A}/2}\nonumber\\
&\times&\mu_{A}^{p}\mu_{B}^{q}\mu_{B}^{r}\mu_{A}^{s}\langle0_{\gamma}|E^{(-)}_{\mathbf{k}',p}(\mathbf{R}_{A})E^{(+)}_{\mathbf{k}',q}(\mathbf{R}_{B})|0_{\gamma}\rangle\langle0_{\gamma}|E^{(-)}_{\mathbf{k}'',r}(\mathbf{R}_{B})E^{(+)}_{\mathbf{k}'',s}(\mathbf{R}_{A})|0_{\gamma}\rangle\Bigr|^{2}\langle0_{\gamma}|E^{(-)}_{\mathbf{k},i}(\mathbf{R}_{A})E^{(+)}_{\mathbf{k},j}(\mathbf{R}_{A})|0_{\gamma}\rangle\nonumber\\
&\simeq&2\left[\frac{\mu_{A}^{p}\mu_{B}^{q}\mu_{B}^{r}\mu_{A}^{s}}{\epsilon^{2}_{0}\hbar^{2}\Gamma_{A}\Delta_{AB}}k_{A}^{4}G^{(0)}_{pq}(\mathbf{R},\omega_{A})G^{*(0)}_{rs}(\mathbf{R},\omega_{A})\right]^{2}.\nonumber
\end{eqnarray}
Both probabilities are $\mathcal{O}(\Gamma^{2}_{A,B}/\Delta^{2}_{AB})$ as anticipated. They must be compensated by interference processes of an order higher than $\mathcal{P}_{1\gamma}^{(d,e)}$ and $\mathcal{P}_{1\gamma}^{(f,g)}$ in virtue of the optical theorem. Note however that for the case of identical atoms the emission probability $\mathcal{P}_{1\gamma}^{(b)}$ is of the same order as that of $\mathcal{P}_{1\gamma}^{(d,e)}$ and $\mathcal{P}_{1\gamma}^{(f,g)}$. Hence, it is in that case the sum $\mathcal{P}_{1\gamma}^{(b)}+\mathcal{P}_{1\gamma}^{(d,e)}+\mathcal{P}_{1\gamma}^{(f,g)}$ that vanishes \cite{BermanRecoil} --the term $\mathcal{P}_{1\gamma}^{(d,e)}$ and part of $\mathcal{P}_{1\gamma}^{(f,g)}$ are however missing in Ref.\cite{BermanRecoil}.

\end{widetext}

\end{document}